\documentclass[journal]{IEEEtran}
\ifCLASSINFOpdf
\else
   \usepackage[dvips]{graphicx}
\fi
\usepackage{url}

\usepackage{graphicx}
\usepackage{amsmath,amsfonts}
\usepackage{array}
\usepackage[caption=false,font=normalsize,labelfont=sf,textfont=sf]{subfig}
\usepackage{algorithmic}
\usepackage{algorithm}
\usepackage{textcomp}
\usepackage{stfloats}
\usepackage{url}
\usepackage{verbatim}
\usepackage{graphicx}
\usepackage{cite}
\usepackage{amsmath,amssymb,amsfonts}
\usepackage{graphicx}
\usepackage{xcolor}
\usepackage{balance}
\usepackage{amsmath}
\usepackage{amsthm}

\newtheorem{Remark}{Remark}
\newenvironment{proov} {{\it \noindent Proof. }} {\hfill $\blacksquare$\par}
\newtheorem{proposition}{Proposition}

\begin{document}
\title{Joint Beamforming for Multi-target Detection and Multi-user Communication in ISAC Systems}
\author{\vspace{0mm} \large{Zongyao Zhao, {\em Graduate Student Member, IEEE}, Zhenyu Liu, {\em Member, IEEE}, Rui Jiang, Zhongyi Li, \\ Xiao-Ping Zhang, {\em Fellow, IEEE}, Xinke Tang, and Yuhan Dong, {\em Senior Member, IEEE}}

\thanks{This work was supported in part by the National Natural Science Foundation of China under Grant 62388102, the GuangDong Basic and Applied Basic Research Foundation under Grant 2022A1515010209, and the Shenzhen Natural Science Foundation under Grant JCYJ20200109143016563. \emph{(Corresponding authors: Xinke Tang and Yuhan Dong)}}

 \thanks{Z. Zhao, Z. Liu, X.-P. Zhang and Y. Dong are with the Shenzhen International Graduate School, Tsinghua University, Shenzhen 518055, China, and also with the Peng Cheng Laboratory, Shenzhen 518055, China (E-mails:
 zhaozong21@mails.tsinghua.edu.cn,  zhenyuliu@sz.tsinghua.edu.cn, xiaoping.zhang@sz.tsinghua.edu.cn, dongyuhan@sz.tsinghua.edu.cn).}

\thanks{Rui Jiang, Zhongyi Li, and X. Tang are with the Peng Cheng Laboratory, Shenzhen 518055, China (E-mail: tangxk@pcl.ac.cn).}

}


\maketitle
\begin{abstract}
Detecting weak targets is one of the main challenges for integrated sensing and communication (ISAC) systems. Sensing and communication suffer from a performance trade-off in ISAC systems. As the communication demand increases, sensing ability, especially weak target detection performance, will inevitably reduce. Traditional approaches fail to address this issue. In this paper, we develop a joint beamforming scheme and formulate it as a max-min problem to maximize the detection probability of the weakest target under the constraint of the signal-to-interference-plus-noise ratio (SINR) of multi-user communication. An alternating optimization (AO) algorithm is developed for solving the complicated non-convex problem to obtain the joint beamformer. The proposed scheme can direct the transmit energy toward the multiple targets properly to ensure robust multi-target detection performance. Numerical results show that the proposed beamforming scheme can effectively increase the detection probability of the weakest target compared to baseline approaches while ensuring communication performance.
\end{abstract}

\begin{IEEEkeywords}
Integrated sensing and communication, joint beamforming, multi-target detection, multi-user communication, weak target detection.
\end{IEEEkeywords}
\IEEEpeerreviewmaketitle

\section{Introduction}
\IEEEPARstart{W}{ith} the development of communication and sensing technology, high-speed communication and high-precision sensing have an increasing demand for spectrum resources\cite{Hassan2016,Liu2022,Zhao2022}. Integrated sensing and communication (ISAC) is an effective way to achieve efficient spectrum utilization. In an ISAC system, wireless communication and wireless sensing adopt a unified waveform, thereby improving energy efficiency, spectrum efficiency, and hardware efficiency at the same time. The ISAC technology is considered to be the key technology to enable these emerging services such as autonomous driving and smart cities\cite{Cui2021,Zhao20241,Zhaoz2024}.

Joint beamforming is important for optimizing the performance of both communication and sensing \cite{Tang2022,Liu2018,Hua2023,LiuX2020,Yuan2021,Zhao2024}. In \cite{Liu2018}, the authors designed a method to jointly optimize multi-user interference (MUI) and radar beampattern. The authors in \cite{Hua2023} proposed a new design scheme aiming to maximize the minimum weighted beampattern gain at the sensing angles of interest under the constraint of signal-to-interference-plus-noise ratio (SINR) for communication. A joint beamforming scheme is proposed in \cite{LiuX2020} to optimize the beampattern under the constraint of communication SINR. The authors in \cite{Zhao2024} proposed a robust beamforming scheme minimizing the largest Cram\'er-Rao bound (CRB) under communication SINR constraint. An information theory method is reported in \cite{Yuan2021}, which jointly optimizes sensing mutual information and communication mutual information. The authors in \cite{Tang2022} studied the detection performance, proposed a design method to maximize the detection probability under the constraint of communication performance, and analyzed the loss of detection performance caused by communication. The authors in \cite{Wang2023} considered the optimization of the Rician target detection performance under MUI constraint.

As can be seen from these previous works, there is a trade-off between the performance of sensing and communication. As the communication demand increases, its detection performance will be degraded. Weak targets are targets with low SINR of echo signals due to small radar cross sections (RCS) or long distances. When weak targets appear in the scene, the detection performance becomes worse. Existing works cannot maintain robust detection performance, especially cannot guarantee the detection performance of weak targets..

In this paper, we design a beamforming scheme for ISAC by optimizing the detection probability of the weakest target while ensuring that the SINR of multi-user communication is above a predefined threshold. The contributions of the paper include the following:
\begin{itemize}
\item We formulate the joint beamforming scheme for ISAC systems as a max-min problem to maximize the detection probability of the weakest target while guaranteeing the SINR of multi-user communication. 
\item We transfer the non-convex max-min problem by introducing auxiliary variable and epigraph equivalent, and develop an alternating optimization (AO) algorithm to solve the problem.
\item We quantify the performance of the designed scheme via simulation. Numerical results show that the proposed scheme can achieve robust detection performance while guaranteeing  communication performance.
\end{itemize}
To the best of our knowledge, this is the first paper considering the jointly optimization of the weakest target detection performance under multi-user communication constraints.

The remainder of this paper is organized as follows. Sec.~\ref{sec2} introduces the system and signal model. We propose the joint beamforming scheme in Sec.~\ref{sec3} and present numerical results in Sec.~\ref{sec4}. Finally, the conclusions are drawn in Sec.~\ref{sec5}.

\emph{Notation}: In this paper, $\mathbb{R}$ and $\mathbb{C}$ represent the real and complex sets respectively. $||\cdot||$ and $|\cdot|_{2}$ are Euclidean norm and absolute
value respectively. $\left( \cdot \right)^T$, $\left( \cdot\right)^*$, and $\left( \cdot \right)^H$ represent the transpose, complex conjugate, and Hermitian transpose, respectively. $\mathrm{Re}\left\{ \cdot\right\}$ returns the real part. $j$ is the imaginary unit, which means $j^2=-1$. $\mathbf{I}_N$ is the $N\times N$ identity matrix. $\mathbf{1}=\left[1,1,\ldots,1\right]^T\in\mathbb{R}^N$. $\mathbf{A}\succeq0$ means that $\mathbf{A}$ is a positive semidefinite matrix. $\otimes$ represents the Kronecker product. $\mathrm{diag}\left(\mathbf{A} \right)$ returns a vector formed by the diagonal elements of matrix $\mathbf{A}$. $\mathrm{tr}\left(\mathbf{A} \right)$ and $\mathrm{rank}\left(\mathbf{A}\right)$ compute the trace and rank of matrix $\mathbf{A}$ respectively. $\mathrm{chol}\left(\mathbf{A}\right)$ returns the Cholesky decomposition of matrix $\mathbf{A}$. $\mathrm{vec}\left(\mathbf{A}\right)$ vectorizes matrix $\mathbf{A}$ by column-stacking. $\min \left\{...\right\}$ returns the smallest element in the set

\section{System and Signal Model} \label{sec2}
\subsection{System Model}
We consider an ISAC system intending to provide downlink communication services for $K$ single-antenna user equipment (UE) and detect $M$ targets simultaneously. We assume that the transmit and receive array of the ISAC system are uniform linear arrays (ULAs) with $N_t$ and $N_r$ antenna elements respectively and $\lambda/2$ inter-spacing between neighboring antenna elements, where $\lambda$ is the carrier wavelength.
The transmitted signal $\mathbf{X}$ can be expressed as
\begin{gather}
\begin{aligned}
\mathbf{X}=\mathbf{W}_C\mathbf{S}_C+\mathbf{W}_R\mathbf{S}_R=\mathbf{WS},
\end{aligned}
\end{gather}
where $\mathbf{W}_C=\left[\mathbf{w}_1,\mathbf{w}_2,\ldots,\mathbf{w}_K \right] \in \mathbb{C} ^{N_t\times K}$ is the communication beamforming matrix, $\mathbf{S}_C=\left[ \mathbf{s}_1,\mathbf{s}_2,\ldots,\mathbf{s}_K \right] ^H\in \mathbb{C} ^{K\times L}$ is communication data
symbol matrix for $K$ UEs with length \emph{L}, $\mathbf{W}_R=\left[\mathbf{w}_{K+1},\mathbf{w}_{K+2},\ldots,\mathbf{w}_{K+N_t} \right] \in \mathbb{C} ^{N_t\times N_t} $ is the sensing beamforming matrix, $\mathbf{S}_R=\left[\mathbf{s}_{K+1},\mathbf{s}_{K+2},\ldots,\mathbf{s}_{K+N_t} \right] ^H\in \mathbb{C} ^{N_t\times L}$ is the dedicated sensing waveform matrix with length \emph{L}, $\mathbf{W}$ is the joint complex beamforming matrix, and $\mathbf{S}$ is the joint data augmentation matrix.

The joint complex beamforming matrix $\mathbf{W}$ and the joint data augmentation matrix $\mathbf{S}$ are given by
\begin{gather}
\begin{aligned}
&\mathbf{W}=\left[ \mathbf{W}_C,\mathbf{W}_R \right] \in \mathbb{C} ^{N_t\times \left( N_t+K \right)},
\label{eq1}
\end{aligned}
\end{gather}
\begin{gather}
\begin{aligned}
&\mathbf{S}=\left[ \begin{array}{c}
	\mathbf{S}_C,
	\mathbf{S}_R\\
\end{array} \right]^{T} \in \mathbb{C} ^{\left( N_t+K \right) \times L}.
\end{aligned}
\end{gather}
When \emph{L} is sufficiently large, the communication data symbol can be regarded to satisfy $\frac{1}{L}\mathbf{S}_C{\mathbf{S}_C}^H=\mathbf{I}_K$ \cite{Liu2018}. At the same time, the sensing waveform $\mathbf{S}_R$ is also carefully designed to satisfy $\frac{1}{L}\mathbf{S}_R{\mathbf{S}_R}^H=\mathbf{I}_{N_t}$. We assume that there is no correlation between sensing waveforms and communication data, which means that $N_t+K$ data streams are uncorrelated and satisfy the following form
\begin{gather}
\begin{aligned}
\frac{1}{L}\mathbf{SS}^H=\mathbf{I}_{N_t+K}.
\end{aligned}
\end{gather}
Therefore, the correlation matrix of waveform $\mathbf{X}$ can be expressed as
\begin{gather}
\begin{aligned}
\mathbf{R}_\mathbf{X}=\frac{1}{L}\mathbf{XX}^H=\mathbf{WW}^{\mathbf{H}}=\mathbf{W}_C\mathbf{W}_{C}^{H}+\mathbf{W}_R\mathbf{W}_{R}^{H}.
\end{aligned}
\end{gather}
\subsection{Target Detection Model}
Assuming that the transmit and receive antenna array are co-located. The received echo signal of targets is given by 
\begin{gather}
\begin{aligned}
\mathbf{Y}_R=\sum_{i=1}^M{\alpha _i\mathbf{b}\left( \theta _i \right) \mathbf{a}^H\left( \theta _i \right)}\mathbf{X}+\mathbf{Z}_R,
\label{eq6}
\end{aligned}
\end{gather}
where $\alpha_i$ and $\theta_i$ are the complex scattering coefficient and azimuth angle of the $i$-th target, respectively. $\mathbf{Z}_{R}$ is the sensing noise matrix with each entry following complex Gaussian distribution $\mathcal{CN}(0,\sigma_R^2)$. Complex scattering coefficient $\alpha_i$ is related to the RCS and the two-way path loss of the target. $\mathbf{a}\left( \theta \right)= \left[ 1,e^{i\pi \sin \theta},e^{2j\pi \sin \theta},...,e^{\left( N_t-1 \right) j\pi \sin \theta} \right] ^T$ and $\mathbf{b}\left( \theta \right)= \left[ 1,e^{i\pi \sin \theta},e^{2j\pi \sin \theta},...,e^{\left( N_r-1 \right) j\pi \sin \theta} \right] ^T$ are steering vectors of transmit and receive antenna array, respectively. We use $\mathbf{b}^H\left( \theta \right)/\left\| \mathbf{b}\left( \theta \right) \right\|$ as the receive filter. Then, the SINR of the $i$-th target echo signal is given by
{\small\begin{gather}
\begin{aligned}
\eta _i=\,\,\frac{|\alpha _i|^2\mathbf{b}^H\left( \theta _i \right) \mathbf{G}\left( \theta _j \right) \mathbf{R}_{\mathbf{x}}\mathbf{G}^H\left( \theta _i \right) \mathbf{b}\left( \theta _i \right) \,\,}{\sum_{j\ne i}{|\alpha _j|^2\mathbf{b}^H\left( \theta _i \right) \mathbf{G}\left( \theta _j \right)}\mathbf{R}_{\mathbf{x}}\mathbf{G}^H\left( \theta _j \right) \mathbf{b}\left( \theta _i \right) +N_r\sigma _{R}^{2}},
\label{eq7}
\vspace{-0.5cm}
\end{aligned}
\end{gather}}
where $\mathbf{G}\left( \theta \right)=\mathbf{b}\left( \theta \right) \mathbf{a}^H\left( \theta \right)$, $\sum_{j\ne i}{|\alpha _j|^2\mathbf{b}^H\left( \theta _i \right) \mathbf{G}\left( \theta _j \right)}\mathbf{R}_{\mathbf{x}}\\\mathbf{G}^H\left( \theta _j \right) \mathbf{b}\left( \theta _i \right)$ is the interference caused by other targets. Then, the detection probability of the $i$-th target is \cite{Tang2022}
\begin{gather}
\begin{aligned}
P_D(i)=\frac{1}{2}\mathrm{erfc}\left\{ \mathrm{erfc}^{-1}\left( 2P_{F} \right) -\sqrt{\eta_i} \right\},
\label{eq8}
\end{aligned}
\end{gather}
where the $\mathrm{erfc}\left( x \right) =\frac{2}{\sqrt{\pi}}\int_x^{\infty}{e^{-t^2}}dt$ is  the complementary error function. $P_{F}$ is the probability of false alarm and is set to be a constant when using constant false-alarm rate (CFAR) detection\cite{Neyman1992,Kay1998}.

\subsection{Multi-user Communication Model}
The signal vector received by the $k$-th UE is given by
\begin{gather}
\begin{aligned}
{\mathbf{y}_{C_k}}=\mathbf{h}_{k}^{H}\mathbf{X}+{\mathbf{z}_{C_k}},
\label{eq9}
\end{aligned}
\end{gather}
where $\mathbf{h}_{k}^{H}$ is the channel from the transmit antenna array to the $k$-th UE. $\mathbf{z}_{C_k}$ is the communication noise vector following $ \mathcal C\mathcal N \left( 0,\sigma_{C}^{2}\mathbf{I}_L \right) $, where $\sigma^2_C$ is the communication noise power. 
The communication SINR of the $k$-th UE is given by
\begin{gather}
\begin{aligned}
\gamma _k=\frac{|\mathbf{h}_{k}^{H}\mathbf{w}_k|^2}{\,\, \sum_{i=1,i\ne k}^{i=K}{|\mathbf{h}_{k}^{H}\mathbf{w}_i|^2}+||\mathbf{h}_{k}^{H}{\mathbf{W}_R}||^2+\sigma _{C}^{2}},
\label{eq10}
\end{aligned}
\end{gather}
where $\sum_{i=1,i\ne k}^{i=K}{|\mathbf{h}_{k}^{H}\mathbf{w}_i|^2}$ is the multi-user interference. $||\mathbf{h}_{k}^{H}\mathbf{W}_R||^2$ is the  the interference caused by the dedicated sensing signal to the communication.

\subsection{Optimization Problem Formulation}\label{AA}
To address the limitation of weak target detection performance in multi-target scenarios, we design a novel beamforming scheme in this section. We formulate a max-min optimization problem maximizing ${P_D}$ of the weakest target to achieve a robust detection performance. The multi-target robust detection design problem under the constraints of communication SINR and power budget can be formulated as follows,
\begin{subequations}\label{eq11}
 \begin{align}
& \mathop{\mathrm{maximize}} \limits_{\,\,\bf{W}}~\mathop{\mathrm{minimize}} \limits_{{i=1,2,\ldots,M}} &&P_{{D}}\left(i \right) 
\\
&~\mathrm{subject~to}   &&\gamma _k\geqslant \gamma_{\mathrm {th}} ~\forall k \label{eq11b}\\
& &&\mathrm{diag}\left( \mathbf{WW}^H \right) =\frac{P_T\mathbf{1}}{N_t}, \label{eq11c}
\end{align}
\end{subequations}
where the joint beamforming matrix $\mathbf{W}$ is the the optimizing variable. For convenience, we assume that different targets have different azimuth angles. $P_T$ is the total transmit power, $\mathrm{diag}\left( \mathbf{WW}^H  \right) =\frac{P_T\mathbf{1}}{N_t}$ is a power budget constraint for each antenna. $\gamma _k\geqslant \gamma_{\mathrm {th}}$ is to set a threshold $\gamma_{\mathrm {th}}$ for the communication SINR of the $K$ UEs to ensure the multi-user communication performance. Problem \eqref{eq11} is a non-convex optimization problem.
\section{Joint Beamforming Scheme} \label{sec3}
It is hard to optimize the $P_{{D}}$ directly. Fortunately, due to \eqref{eq12}, the function of $P_{{D}}$ with respect to the target echo SINR is monotonic so that maximizing  $P_{{D}}\left( i \right)$ is equivalent to maximizing $\eta_i$. We can transform \eqref{eq11} into the following form
\begin{subequations}\label{eq12}
 \begin{align}
& \mathop{\mathrm{maximize}} \limits_{\,\,\bf{W}}~\mathop{\mathrm{minimize}} \limits_{{i=1,2,\ldots,M}} &&\eta_i 
\\
&~\mathrm{subject~to}   &&\eqref{eq11b}~ \eqref{eq11c} \nonumber
\end{align}
\end{subequations}
Introducing the auxiliary variables $q$ and $\mathbf{R}_\mathbf{X}= \frac{1}{L}\mathbf{XX}^H=\mathbf{WW}^{H}$, $\mathbf{W}_k=\mathbf{w}_k\mathbf{w}_{k}^{H}$, and $\mathbf{Q}_k=\mathbf{h}_{k}\mathbf{h}_{k}^{H}$, the epigraph equivalent \cite{boyed2004} form can be written as
\begin{subequations}\label{eq13}
\begin{align}
& \mathop{\mathrm{minimize}} \limits_{\left\{\mathbf{W}_k \right\}_{k=1,2,\ldots,K},\mathbf{R}_\mathbf{X},q} &&-q 
\\
&~\mathrm{subject~to}  && \eta_i-q\geqslant 0,~ \forall i \label{eq13b}\\
& && q\geqslant 0\label{eq13c}\\
& &&\mathbf{W}_k\succeq 0,~\forall k\\
& &&\mathbf{R}_\mathbf{X}-\sum_{k=1}^K\mathbf{W}_k\succeq 0\\
& &&\mathrm{rank}\left( \mathbf{W}_k \right) =1, ~\forall k \label{eq13f}\\
& &&\eqref{eq11b}~ \eqref{eq11c}. \nonumber
\end{align}
\end{subequations}
Substitute \eqref{eq10} into \eqref{eq11b}, it can be reformulated as 
\begin{gather}
\begin{aligned}
\mathrm{tr}\left( \mathbf{Q}_k\mathbf{W}_k \right) -\gamma_{\mathrm {th}} \mathrm{tr}\left( \mathbf{Q}_k\left( \mathbf{R}_\mathbf{X}-\mathbf{W}_k \right) \right)\geqslant \gamma_{\mathrm {th}} \sigma _{C}^{2},~\forall k,
\label{eq14}
\end{aligned}
\end{gather}

\begin{figure*}[t]

\begin{align}\label{eq15}
 |\alpha _i|^2\mathbf{b}^H\left( \theta _i \right) \mathbf{G}\left( \theta _j \right) \mathbf{R}_{\mathbf{x}}\mathbf{G}^H\left( \theta _i \right) \mathbf{b}^H\left( \theta _i \right) -y\sum_{j\ne i}{|\alpha _j|^2\mathbf{b}^H\left( \theta _i \right) \mathbf{G}\left( \theta _j \right)}\mathbf{R}_{\mathbf{x}}\mathbf{G}^H\left( \theta _j \right) \mathbf{b}^H\left( \theta _i \right) +N_r\sigma _{R}^{2}-q\geqslant 0,~\forall i,
\end{align}
\hrulefill
\vspace{-0.6em}
\end{figure*}
\setcounter{equation}{19}
\begin{figure*}[t]
\begin{align}\label{eq16}
y^{\left( t+1 \right)}=\min \left\{ y_{i}^{*}~|~i=1,2,...,M \right\},~~ y_{i}^{*}=\frac{|\alpha _i|^2\mathbf{b}^H\left( \theta _i \right) \mathbf{G}\left( \theta _i \right) \mathbf{R}_{\mathbf{x}}^{(t)}\mathbf{G}^H\left( \theta _i \right) \mathbf{b}^H\left( \theta _i \right)}{\sum_{j\ne i}{|\alpha _j|^2\mathbf{b}^H\left( \theta _i \right) \mathbf{G}\left( \theta _j \right)}\mathbf{R}_{\mathbf{x}}^{(t)}\mathbf{G}^H\left( \theta _j \right) \mathbf{b}^H\left( \theta _i \right) +N_r\sigma _{R}^{2}}
\end{align}
\hrulefill
\vspace{-0.6em}
\end{figure*}
\setcounter{equation}{15}
By introducing an auxiliary variables $y$, which represents minimum echo signal SINR of all targets, we can convert the constraint \eqref{eq13b} into \eqref{eq15}. The optimization problem can be transformed as
\begin{subequations}\label{eq17}
 \begin{align}
& \mathop{\mathrm{minimize}} \limits_{\left\{\mathbf{W}_k \right\}_{k=1,2,\ldots,K},\mathbf{R}_\mathbf{X},q,y} &&-q 
\\
&~\mathrm{subject~to}  && \eqref{eq11c}, \eqref{eq13c}-\eqref{eq13f}, \eqref{eq14}, \eqref{eq15}. \nonumber
\end{align}
\end{subequations}

In order to solve this problem, we design an AO based algorithm. 

\textbf{Update $\mathbf{W}$}: At the $t$-th step, given $y=y^{(t)}$, the optimization problem can be expressed as 
\begin{subequations}\label{eq18}
 \begin{align}
& \mathop{\mathrm{minimize}} \limits_{\left\{\mathbf{W}_k \right\}_{k=1,2,\ldots,K},\mathbf{R}_\mathbf{X},q} &&-q 
\\
&~\mathrm{subject~to}  &&\eqref{eq11c}, \eqref{eq13c}-\eqref{eq13f}, \eqref{eq14}, \eqref{eq15}. \nonumber
\end{align}
\end{subequations}
The above problem is still a non-convex problem due to the rank-1 constraint \eqref{eq13f}. However, we can use the a semidefinite relaxation (SDR) method to obtain the optimal solution. First, we ignore the rank-1 constraint \eqref{eq13f} and obtain the optimal solution $\overline{q}$, $\overline{\mathbf{R}}_\mathbf{X}$, and $\overline{\mathbf{W}}_k$. Next, we calculate the communication and sensing beamforming matrix $\mathbf{W}_C$ via the following formulas \cite{LiuX2020}
\begin{gather}
\begin{aligned}
\label{eq19}
&\widetilde{\mathbf{W}}_k=\frac{\overline{\mathbf{W}}_k\mathbf{Q}_k\overline{\mathbf{W}}_{k}^{H}}{\mathrm{tr}\left( \mathbf{Q}_k\overline{\mathbf{W}}_k \right)},\,\,\forall k\leqslant K\\
&\overline{\mathbf{w}}_k=\left( \mathbf{h}_{k}^{H}\widetilde{\mathbf{W}}_k\mathbf{h}_k \right) ^{-1/2}\widetilde{\mathbf{W}}_k\mathbf{h}_k,\,\,\forall k\leqslant K\\
&\mathbf{W}_C=\left[\overline{\mathbf{w}}_1,\overline{\mathbf{w}}_2,\ldots,\overline{\mathbf{w}}_K \right]. \\
&\mathbf{W}_R=\mathrm{chol}\left(\overline{\mathbf{R}}_\mathbf{X}-\sum_{k=1}^K\widetilde{\mathbf{W}}_k\right).
\end{aligned}
\end{gather}
According to \eqref{eq1}, the joint beamforming matrix $\mathbf{W}^{(t)}$ can finally be computed as 
\begin{gather}
\begin{aligned}
\label{eq20}
\mathbf{W}^{(t)}=\left[ \mathbf{W}_C,\mathbf{W}_R \right].
\end{aligned}
\end{gather}
\setcounter{equation}{20}
\textbf{Update $y$}: At $t$ step, given $\mathbf{W}=\mathbf{W}^{(t)}$, we have $\mathbf{R}_{\mathbf{X}}^{(t)}=\mathbf{W}\mathbf{W}^{H}$. Then $y$ can be updated by \eqref{eq16}. 

We can update $y$ and $\mathbf{W}$ iteratively to obtain the joint beamformer.  We summarize the proposed AO algorithm in Algorithm \ref{alg1}. 

\renewcommand{\algorithmicrequire}{ \textbf{Input:}} 
\renewcommand{\algorithmicensure}{ \textbf{Output:}} 
\begin{algorithm}[!t]
    \caption{AO algorithm for \eqref{eq17}}
    \label{alg1}
  \begin{algorithmic} [1]
 \REQUIRE \, 
    $P_T$, $M$, $K$, $N_t$, $N_r$, $\sigma_{C}^2$, $\sigma_{R}^2$, $\gamma_\mathrm{th}$, $\{\mathbf{h}_{k}~|~k=1,2,...,K\}$, $\{\alpha_{i}~|~i=1,2,...,M\}$
    \ENSURE Designed beamforming matrix $\mathbf{W}^{*}$.
    \renewcommand{\algorithmicensure}{ \textbf{Steps:}}
    \ENSURE \, 
   \STATE Initialize: $\delta_\mathrm{th}$, ${T}_{\text{max}}$ , $t=1$, 
    $\delta=\infty$, $y^{(1)}$. 
    
    \STATE \textbf{while} $t \leqslant {T}_{\text{max}}$ and  $\delta \geqslant \delta_\mathrm{th}$  \textbf{do}
    \STATE ~~~~~Obtain $\overline{q}$, $\overline{\mathbf{R}}_\mathbf{X}$, and $\overline{\mathbf{W}}_k$ by solving \eqref{eq18} without rank-1 constraint \eqref{eq13f}.
    \STATE ~~~~~Update $\mathbf{W}^{(t)}$ by using \eqref{eq19}\eqref{eq20}.
    \STATE ~~~~~Update $y^{t+1}$ by \eqref{eq16}. 
    \STATE ~~~~ $\delta=y^{(t+1)}-y^{(t)}$.
    \STATE ~~~~~$t=t+1$.
    \STATE \textbf{end while}
    \STATE Return $\mathbf{W}^{{*}}=\mathbf{W}$.
\end{algorithmic}
\end{algorithm}

Next, we analyze the convergence of the proposed algorithm. We give the following proposition.
\begin{proposition}
Minimum echo signal SINR of all targets $y$ increases with iterations of the proposed algorithm, i.e.,
\begin{align}
y^{\left( t+1 \right)}\geqslant y^{\left( t \right)}
\end{align}
\label{lm1}
\end{proposition}
\vspace{-0.5cm}
\begin{proov}
See Appendix~\ref{AP1}.
\end{proov}
\begin{Remark}
Note that $y$ is the minimum echo signal SINR of all targets. Since the signal power is not infinite, there is always an upper bound to the SINR. Hence, the proposed alternating optimizing algorithm converges.
\end{Remark}

Now, we analyze the complexity of the proposed algorithm. Problem \eqref{eq18} without rank-1 constraint is a semidefinite program (SDP)\cite{boyed2004}. Hence, given a solution accuracy $\varepsilon$, the worst-case complexity to solve the SDP problem with the primal-dual interior-point method is about $\mathcal{O} \left(K^{4.5}N_{t}^{4.5}\log \left( 1/\varepsilon \right) \right)$\cite{Luo2010,Hua2023}. Assuming that $I_{iter}$ of iterations are required for the algorithm to converge, then the total complexity is $\mathcal{O} \left( I_{iter}K^{4.5}N_{t}^{4.5}\log \left( 1/\varepsilon \right) \right)$.

\section{Simulation Results} \label{sec4}
\begin{table}[b]
	\centering
	\fontsize{9.0}{8.0}\selectfont
	 \caption{Parameters in Simulations.}\label{tab1}

		{\begin{tabular}{p{1.0cm} p{1.0cm} p{1.0cm} p{1.0cm}l p{1.0cm} p{1.0cm}}
				\hline 
                    \hline
                    
				Parameter & Value &Parameter & Value &Parameter & Value \\
				 \hline
			$N_t$ & $16$ &	$N_r$ &$16$ &$M$ & $3$ \\				
				$K$	& $6$ & $P_T$	& $30~\mathrm{dBm}$  &  $\sigma_{R}^2$	& $0~\mathrm{dBm}$ \\
				$\sigma_{C}^2$ & $0~\mathrm{dBm}$ &	$\gamma_{\mathsf{th}}$ & $15~\mathrm{dB}$	& $\delta_{\mathsf{th}}$  & $10^{-5}$  \\				$\left|\alpha_1\right|^2$ & $0.25$ &	$\left|\alpha_2\right|^2$ & $0.0001$	& $\left|\alpha_3\right|^2$  & $0.04$  \\	$\theta_1$ & $-45^{\circ}$ &	$\theta_2$ & $0^{\circ}$	& $\theta_3$  & $60^{\circ}$  \\	
    
				\hline 
		\end{tabular}}
\end{table}
In this section, we present the numerical results of the proposed beamforming scheme. In order to examine the detection robustness of the proposed scheme, the scattering strengths of targets are set differently as $\left|\alpha_1\right|^2=0.25$, $\left|\alpha_2\right|^2=0.0001$, and $\left|\alpha_3\right|^2=0.04$. The angle of the three targets are set as $\theta_1=-45^{\circ }$, $\theta_2=0^{\circ}$, and $\theta_3=60^{\circ}$. Obviously, target-2 is the weakest target. We set the transmit and receive array with the same element number, i.e., $N_t =N_r=16$, and set the transmit power as $P_T=30$ dBm and the noise power of communication and sensing as $\sigma _{C}^{2}=\sigma _{R}^{2}=0$ dBm. We assume Rayleigh fading channel with channel matrix in which all elements are independent complex Gaussian variables with zero mean and unit variance. All simulation parameters are listed in Table \ref{tab1}.

We use several baseline methods to compare with the proposed scheme (Joint Scheme). Max-min weighted Beampattern-1 and Max-min weighted Beampattern-2 are the schemes proposed in \cite{Hua2023} with weight $[1,1,1]$ and $[1/\left|\alpha_1\right|^2,1/\left|\alpha_2\right|^2,1/\left|\alpha_3\right|^2]$, respectively. Beampattern Approx. is the scheme proposed in \cite{LiuX2020}. Robust CRB is the scheme proposed in \cite{Zhao2024}. Omini Strict is the full orthogonal beamforming scheme with $\mathbf{R}_\mathbf{X}=\frac{P_T}{N_t}\mathbf{I}_{N_t}$, which radiates the same energy at different angles. Sensing Only (Proposed) is the scheme produced by solving the problem \eqref{eq12} without the communication constraint \eqref{eq11c}, which means the scheme only considers the sensing performance.
\begin{figure}[!t]
\captionsetup{font=small}
\begin{center}
\includegraphics[width=0.45\textwidth]{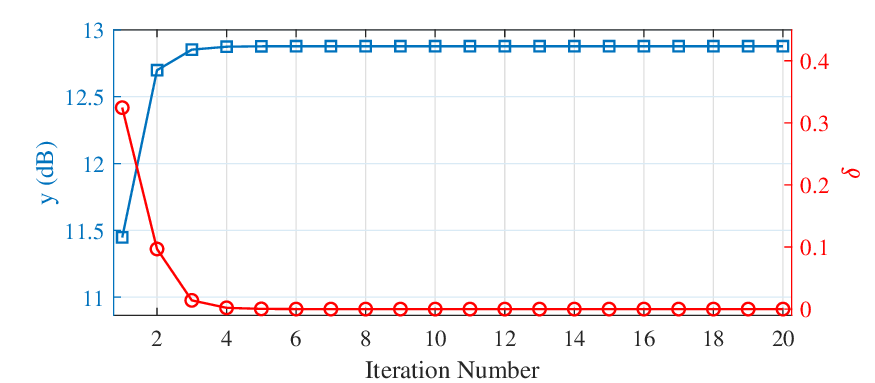}
\end{center}
\vspace{-0.5cm}
\caption{Convergence performance of the proposed Algorithm 1.}
\label{Fig1}
\vspace{-0.5cm}
\end{figure}

We first verified the convergence of the proposed AO algorithm. Fig.~\ref{Fig1} shows the minimum SINR $y$ of targets and increased value $\delta$  versus the number of iterations. It can be seen that the proposed algorithm fully converges within 5 iterations, indicating that it has fast convergence performance.

\begin{figure}[!t]
\captionsetup{font=small}
\begin{center}
\includegraphics[width=0.45\textwidth]{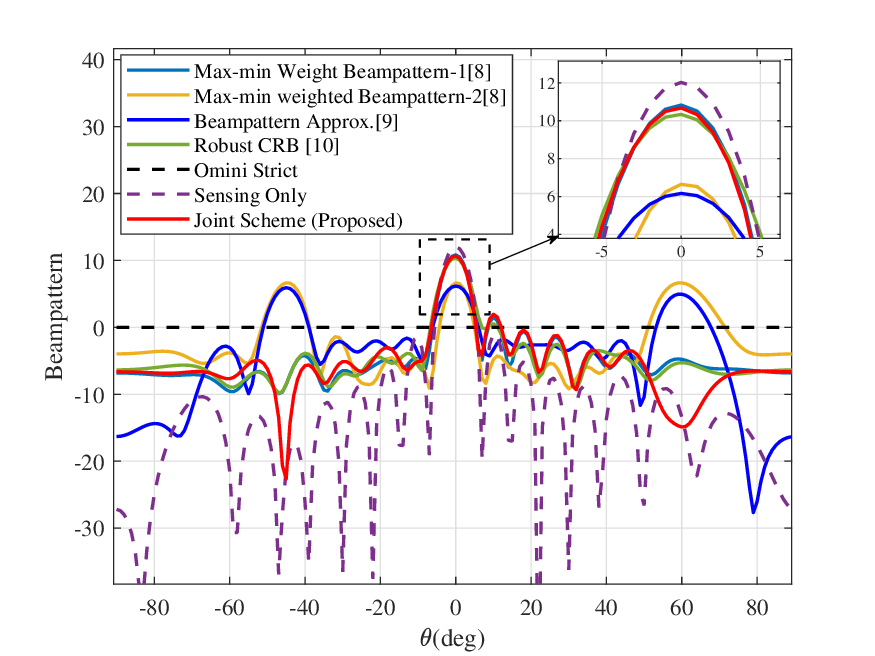}
\end{center}
\vspace{-0.5cm}
\caption{Beampatterns of various beamforming schemes.}
\label{Fig2}
\vspace{-0.5cm}
\end{figure}

Fig.~\ref{Fig2} shows the beampattern of various schemes with $K=6$, and $\gamma_{\mathrm {th}}=15$ dB. From Fig.~\ref{Fig2} we can see that the proposed scheme can provide more energy for the weak target with the improved detection performance. At the same time, it also reduces the energy for the strong target to avoid excessive side lobe energy interfering with weak target detection.
\begin{figure}[!t]
\captionsetup{font=small}
\begin{center}
\includegraphics[width=0.45\textwidth]{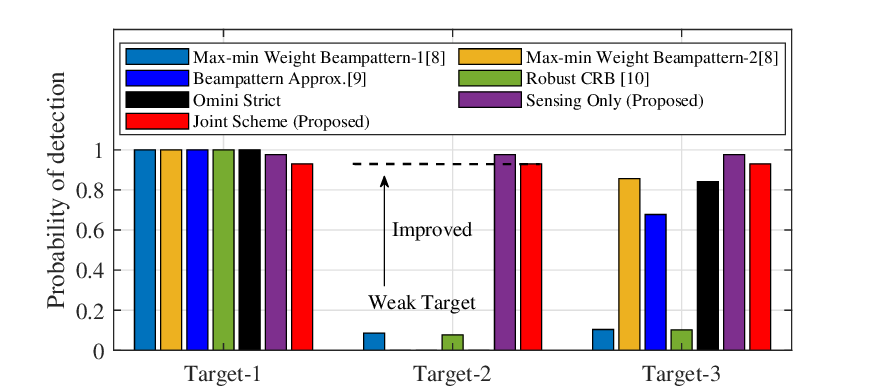}
\end{center}
\vspace{-0.5cm}
\caption{${P_{D}}$ of three targets for various beamforming schemes.}
\label{Fig3}
\vspace{-0.6cm}
\end{figure}

Fig.~\ref{Fig3} shows the ${P_{D}}$ of the three targets for these beamforming methods. The proposed scheme shows a higher ${P_{D}}$ for the Target-2 and Target-3. This shows that the system's ability to detect weak targets has been enhanced.

\begin{figure}[!t]
\captionsetup{font=small}
\begin{center}
\includegraphics[width=0.45\textwidth]{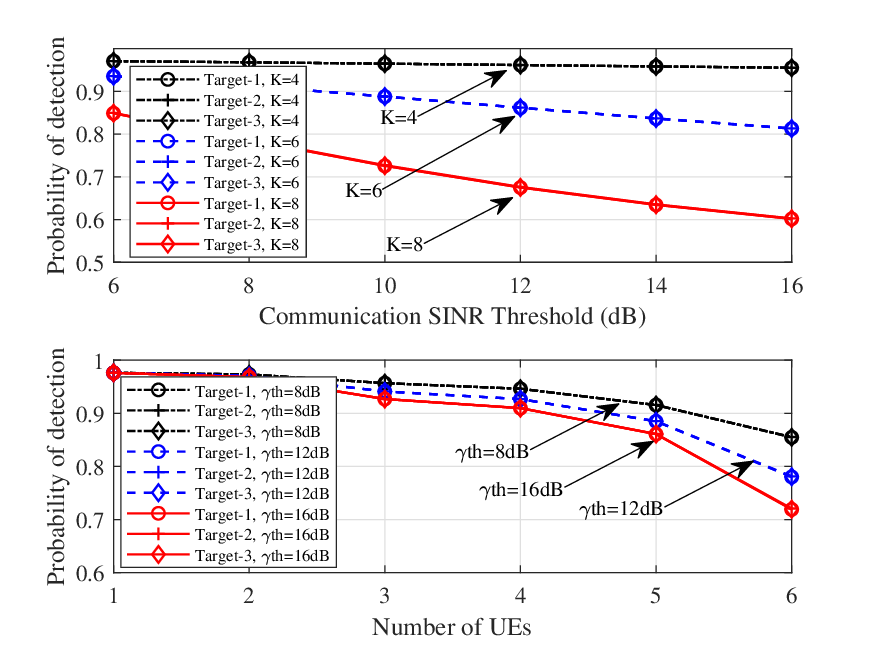}
\end{center}
\vspace{-0.6cm}
\caption{${P_{D}}$ of the proposed scheme versus the communication SINR threshold and the different number of UEs $K$.}
\label{Fig4}
\vspace{-0.5cm}
\end{figure}

We have also examined how the ${P_{D}}$ varies with the communication SINR threshold and the number of UEs. Fig.~\ref{Fig4} shows how the ${P_{D}}$ varies with respect to the communication threshold $\gamma_{\mathrm {th}}$ and the number of UEs, i.e., $K$. From Fig.~\ref{Fig4}, we can observe the performance trade-off between communication and sensing: the detection probability ${P_{D}}$ decreases as the communication SINR threshold $\gamma_{\mathrm {th}}$ and the number of communication UEs $K$ increase.

In general, numerical results show that the proposed scheme can achieve robust multi-target detection performance while guaranteeing communication performance. The proposed scheme can facilitate V2X ISAC applications, especially sensing important but weak targets such as pedestrians with poor scattering capabilities.

\section{Conclusion} \label{sec5}
In this paper, we proposed a joint beamforming scheme for ISAC systems that simultaneously supports multi-target detection and multi-user communication. The proposed scheme can maximize the weakest target detection probability while guaranteeing that the SINR of multi-user communication is above a predefined threshold. An AO optimization algorithm is developed to solve the complex non-convex problem corresponding to the proposed scheme. The effectiveness of the proposed scheme has been proven through a series of simulations. Compared with other baseline schemes, the proposed scheme has stronger detection performance for weak targets. 

\appendices
\section{Proof of Proposition 1}
\label{AP1}
At the $t$ iteration step, we have $\mathbf{R}_{\mathbf{X}}^{(t)}=\mathbf{W}^{(t)}\mathbf{W}^{(t)H}$ is the optimal solution of \eqref{eq18}. Therefore, $\mathbf{R}_{\mathbf{X}}^{(t)}$ satisfies the constraint \eqref{eq15} of the problem \eqref{eq18}. Substitute $\mathbf{R}_{\mathbf{X}}^{(t)}$ in \eqref{eq15} and divide both sides of \eqref{eq15} by $\sum_{j\ne i}{|\alpha _j|^2\mathbf{b}^H\left( \theta _i \right) \mathbf{G}\left( \theta _j \right)}\mathbf{R}_{\mathbf{x}}\mathbf{G}^H\\\left( \theta _j \right) \mathbf{b}^H\left( \theta _i \right)+N_r\sigma _{R}^{2}$, we have 
{\small
\begin{align}
\label{eq22}
&\frac{|\alpha _i|^2\mathbf{b}^H\left( \theta _i \right) \mathbf{G}\left( \theta _j \right) \mathbf{R}_{\mathbf{X}}^{\left( t \right)}\mathbf{G}^H\left( \theta _i \right) \mathbf{b}^H\left( \theta _i \right)}{\sum_{j\ne i}{|\alpha _j|^2\mathbf{b}^H\left( \theta _i \right) \mathbf{G}\left( \theta _j \right)}\mathbf{R}_{\mathbf{X}}^{\left( t \right)}\mathbf{G}^H\left( \theta _j \right) \mathbf{b}^H\left( \theta _i \right) +N_r\sigma _{R}^{2}}-y^{\left( t \right)}\nonumber\\
&\geqslant \frac{q}{\sum_{j\ne i}{|\alpha _j|^2\mathbf{b}^H\left( \theta _i \right) \mathbf{G}\left( \theta _j \right)}\mathbf{R}_{\mathbf{X}}^{\left( t \right)}\mathbf{G}^H\left( \theta _j \right) \mathbf{b}^H\left( \theta _i \right) +N_r\sigma _{R}^{2}}
\end{align} }
From the constraint \eqref{eq13c} of \eqref{eq13}, we can know that $q\geqslant0$. Therefore, the right hand side of \eqref{eq22} is larger than 0, i.e., 
 {\small\begin{align}
\label{eq23}
\frac{q}{\sum_{j\ne i}{|\alpha _j|^2\mathbf{b}^H\left( \theta _i \right) \mathbf{G}\left( \theta _j \right)}\mathbf{R}_{\mathbf{X}}^{\left( t \right)}\mathbf{G}^H\left( \theta _j \right) \mathbf{b}^H\left( \theta _i \right) +N_r\sigma _{R}^{2}}\geqslant0.
\end{align}}
According to \eqref{eq16}, the first part of left hand side of \eqref{eq22} is equal to $y_{i}^{*}$. So, we can get 
\begin{align}
\label{eq24}
y_{i}^{*}-y^{(t)}\geqslant 0,\forall i,
\end{align}
The $y^{\left( t+1 \right)}=\min \left\{ y_{i}^{*}~|~i=1,2,...,M \right\}$, so that $y^{(t+1)}-y^{(t)}\geqslant 0$, which complete the proof.


\begin{thebibliography}{00}
\bibitem{Hassan2016} A. Hassanien, M. G. Amin, Y. D. Zhang, and F. Ahmad, ``Signaling strategies for dual-function radar communications: An overview,'' {\em IEEE Aerosp. Electron. Syst. Mag.}, vol. 31, no. 10, pp. 36--45, Oct. 2016.

\bibitem{Zhao2022} Z. Zhao, X. Tang, and Y. Dong, ``Cognitive waveform design for dual-functional MIMO radar-communication systems,'' in {\em Proc. IEEE Global Commun. Conf. (GLOBECOM)}, Dec. 2022, pp. 5607--5612.

\bibitem{Liu2022} F. Liu, Y. Cui, C. Masouros, J. Xu, T. X. Han, Y. C. Eldar, and S. Buzzi, ``Integrated sensing and communications: Towards dual-functional wireless networks for 6G and beyond,'' {\em IEEE J. Sel. Areas Commun.}, vol. 40, no. 6, pp. 1728--1767, Jun. 2022.

\bibitem{Cui2021} Y. Cui, F. Liu, X. Jing and J. Mu, ``Integrating sensing and communications for ubiquitous IoT: Applications, trends, and challenges,'' {\em IEEE Network}, vol. 35, no. 5, pp. 158-167, Nov. 2021.

\bibitem{Zhao20241} Z. Zhao, T. Wei, Z. Liu, X. Tang, X.-P. Zhang, and Y. Dong, ``Joint Beamforming for Backscatter Integrated Sensing and Communication,'' {\em in Proc. IEEE Global Commun. Conf. (GLOBECOM)}, Dec. 2024, arXiv:2409.0279.

\bibitem{Zhaoz2024} Z. Zhao, Y. Dong, T. Wei, X.-P. Zhang, X. Tang, and Z. Liu ``B-ISAC: Backscatter integrated sensing and communication for 6G IoE applications,'' arXiv:2407.19235.

\bibitem{Liu2018} F. Liu, L. Zhou, C. Masouros, A. Li, W. Luo, and A. Petropulu, ``Toward dual-functional radar-communication systems: Optimal waveform design,'' {\em IEEE Trans. Signal Process.}, vol. 66, no. 16, pp. 4264--4279, Aug. 2018.

\bibitem{Hua2023} H. Hua, J. Xu and T. X. Han, ``Optimal transmit beamforming for integrated sensing and communication,'' {\em IEEE Trans. Veh. Technol.}, vol. 72, no. 8, pp. 10588-10603, Aug. 2023.

\bibitem{LiuX2020} X. Liu, T. Huang, N. Shlezinger, Y. Liu, J. Zhou, and Y. C. Eldar, ``Joint transmit beamforming for multiuser MIMO communications and MIMO radar,'' {\em IEEE Trans. Signal Process.}, vol. 68, pp. 3929--3944, Jun. 2020.

\bibitem{Zhao2024} Z. Zhao, {\em et al.}, ``Joint beamforming scheme for ISAC systems via robust Cramér–Rao bound optimization,'' {\em IEEE Wireless Commun. Lett.}, vol. 13, no. 3,  pp. 889--893, Jan. 2024.

\bibitem{Yuan2021} X. Yuan {\em et al}., ``Spatio-temporal power optimization for MIMO joint communication and radio sensing systems with training overhead,'' {\em IEEE Trans. Veh. Technol.}, vol. 70, no. 1, pp. 514-528, Jan. 2021.

\bibitem{Tang2022} B. Tang and P. Stoica, ``MIMO multifunction RF systems: Detection performance and waveform design,'' {\em IEEE Trans. Signal Process.}, vol. 70, pp. 4381-4394, Aug. 2022.

\bibitem{Wang2023} X. Wang, B. Tang, W. Wu and D. Li., ``Relative entropy-based waveform optimization for Rician target detection with dual-function radar communication systems,'' {\em IEEE Sens. J.}, vol. 23, no. 10, pp. 10718-10730, May 2023.


\bibitem{Fortu2020} S. Fortunati, L. Sanguinetti, F. Gini, M. S. Greco, and B. Himed, ``Massive MIMO radar for target detection,'' {\em IEEE Trans. Signal Process.}, vol. 68, pp. 859-871, Jan. 2020.

\bibitem{Neyman1992} J. Neyman and E. S. Pearson, ``On the problem of the most efficient tests of statistical hypotheses,'' {\em Philosophical Trans. the Royal Society of London}, vol. 231, no. 694-706, pp. 289--337, 1992.

\bibitem{Kay1998}S. M. Kay, {\em Fundamentals of Statistical Signal Processing-Volume II: Detection Theory}, Englewood Cliffs, NJ, USA: Prentice Hall, 1998.

\bibitem{boyed2004}S.Boyd and L. Vandenberghe, {\em Convex Optimization.}, Cambridge University Press, 2004.

\bibitem{Luo2010} Z.-Q. Luo, W.-K. Ma, A. M.-C. So, Y. Ye, and S. Zhang, ``Semidefinite relaxation of quadratic optimization problems,'' {\em IEEE Signal Process. Mag.}, vol. 27, no. 3, pp. 20--34, May 2010.

\bibliographystyle{IEEEtran}
\end{thebibliography}
\end{document}